\documentclass[12pt]{iopart}
\usepackage{graphicx}
\usepackage{xcolor}
\usepackage{iopams}  
\usepackage{hyperref}
\usepackage{braket}

\usepackage{caption}
\expandafter\let\csname equation*\endcsname\relax 
\expandafter\let\csname endequation*\endcsname\relax
\usepackage{amsmath}

\DeclareMathOperator*{\argmax}{arg\,max}
\newcommand{\boldtheta}{\textbf{\boldmath$\theta$}}

\begin{document}

\title{Preparation of ordered states in ultra-cold gases using Bayesian optimization}

\author{Rick Mukherjee\textsuperscript{1}, Fr\'ed\'eric Sauvage\textsuperscript{1}, Harry Xie\textsuperscript{1}, Robert L\"{o}w\textsuperscript{2} and Florian Mintert\textsuperscript{1}}

\address{$^1$ Department of Physics, Imperial College London, SW7 2AZ, London, UK}
\address{$^2$ 5. Physikalisches Institut and Center for Integrated Quantum Science and Technology, Universit\"{a}t Stuttgart, Pfaffenwaldring 57, 70569 Stuttgart, Germany}

\ead{r.mukherjee@imperial.ac.uk}
\vspace{10pt}
\begin{indented}
\item[]September 2019
\end{indented}

\begin{abstract}
Ultra-cold atomic gases are unique in terms of the degree of controllability, both for internal and external degrees of freedom. This makes it possible to use them for the study of complex quantum many-body phenomena. However in many scenarios, the prerequisite condition of faithfully preparing a desired quantum state despite decoherence and system imperfections is not always adequately met. To pave the way to a specific target state, we implement quantum optimal control based on Bayesian optimization. The probabilistic modeling and broad exploration aspects of Bayesian optimization are particularly suitable for quantum experiments where data acquisition can be expensive. Using numerical simulations for the superfluid to Mott-insulator transition for bosons in a lattice as well as for the formation of Rydberg crystals as explicit examples, we demonstrate that Bayesian optimization is capable of finding better control solutions with regards to finite and noisy data compared to existing methods of optimal control.
\end{abstract}

\section{Introduction}
In this paper, the focus is on creating spatially ordered states in two different ultra-cold systems, atoms in optical lattice and highly excited Rydberg atoms, as testbeds for Bayesian optimization. Optical lattices \cite{UCA2012,UCA2017} provide a natural and versatile platform to study strongly correlated condensed matter systems \cite{Qsim1, Qsim2}, with implementations in quantum information processing \cite{Deutschnew}  and quantum metrology \cite{Qmet1,Qmet2,Qmet3}. Likewise the extraordinary properties of Rydberg atoms \cite{Gallagher_1988} in combination with ultra-cold gases establish a facile way to realize strongly interacting many-body systems \cite{Robicheaux, Honer, Zeiher2}. Recently Rydberg atoms trapped in tweezers have been used to produce some of the largest non-classical states \cite{Omran570} and have emerged as a serious competitor for the realization of quantum computer \cite{Saffman_rev}. While adiabatic preparation still remains an intuitive and straightforward way to create these interesting states, it requires long timescales especially for large systems with diminishing energy gaps. This makes it prone to accumulation of errors due to prolonged interaction with the environment. Furthermore, there is no guarantee that a particular target state of interest is adiabatically connected to the initial state. Thus in recent years, there has been a growing interest in applying optimal control to ultra-cold systems \cite{ Omran570, CRAB2,van2016optimal, wigley2016fast, DRLTRAP}.

At the core of any quantum optimal control framework, lie optimization algorithms that aim at maximizing the figure of merit of the experiment with respect to input control fields.  A figure of merit (FoM) measures the success of the experiment in achieving a specific task. This could be the preparation of a highly desirable quantum state, implementation of a gate or even achieving a certain phase transition. The input control fields of a typical quantum experiment can be tunable variables such as external electric or magnetic fields, amplitude and phase of lasers or microwave pulses. The explicit dependence of the FoM on these input control fields defines the control (optimization) landscape. To obtain an optimal solution, an optimization scheme generally requires to probe several points in the control landscape. Undoubtedly the more points the optimization algorithm can probe the better is the optimal solution. However, if the FoM is obtained from a quantum experiment, probing several points in the control landscape would imply repeating the experiment several times. Depending on the complexity of the quantum task at hand and the optimization routine, the experiment may have to be performed for unusually high number of times, which is highly inefficient. This constraint applies equally well to numerical calculations where the Hilbert space grows exponentially with system size. Most conventional methods of optimization find fast protocols but not necessarily efficiently with regards to the number of data points needed from the control landscape. Furthermore, there is always some uncertainty in evaluating the FoM due to experimental noise and imperfections which makes the optimization process even more challenging.

In this regard, Bayesian Optimization (BO) \cite{brochu2010tutorial,snoek2012practical,shahriari2015taking} is an attractive option and has been successfully used in robotics \cite{cully2015robots,calandra2016bayesian}, machine-learning \cite{snoek2012practical} and has recently made its way to quantum optimal control problems \cite{wigley2016fast, zhu2018training, Henson2018,Nakamura2019}. Section \ref{CLQOLS} discusses how BO fits into the general framework of quantum optimal control followed by a discussion outlining the essential steps of BO in Section~\ref{BOSection}. In Section \ref{orderorder}, BO is implemented for two different examples. In the first example, BO's efficiency with regards to the minimum number of evaluations required to drive a superfluid (SF) to Mott insulator (MI) state transition is established when compared to several other optimization routines. In the second example, BO is applied to a strongly interacting gas of trapped Rydberg atoms. Although initial studies of Rydberg gases revealed the possibility of creating ordered structures via a phase transition \cite{Weimer}, the presence of a quantum critical point makes the realization of crystalline ordering with large number of Rydberg excitations still challenging. Section \ref{final} summarizes these results and provides an outlook for future work.

\begin{figure}[t!]
\centering
\includegraphics[width=0.95\columnwidth]{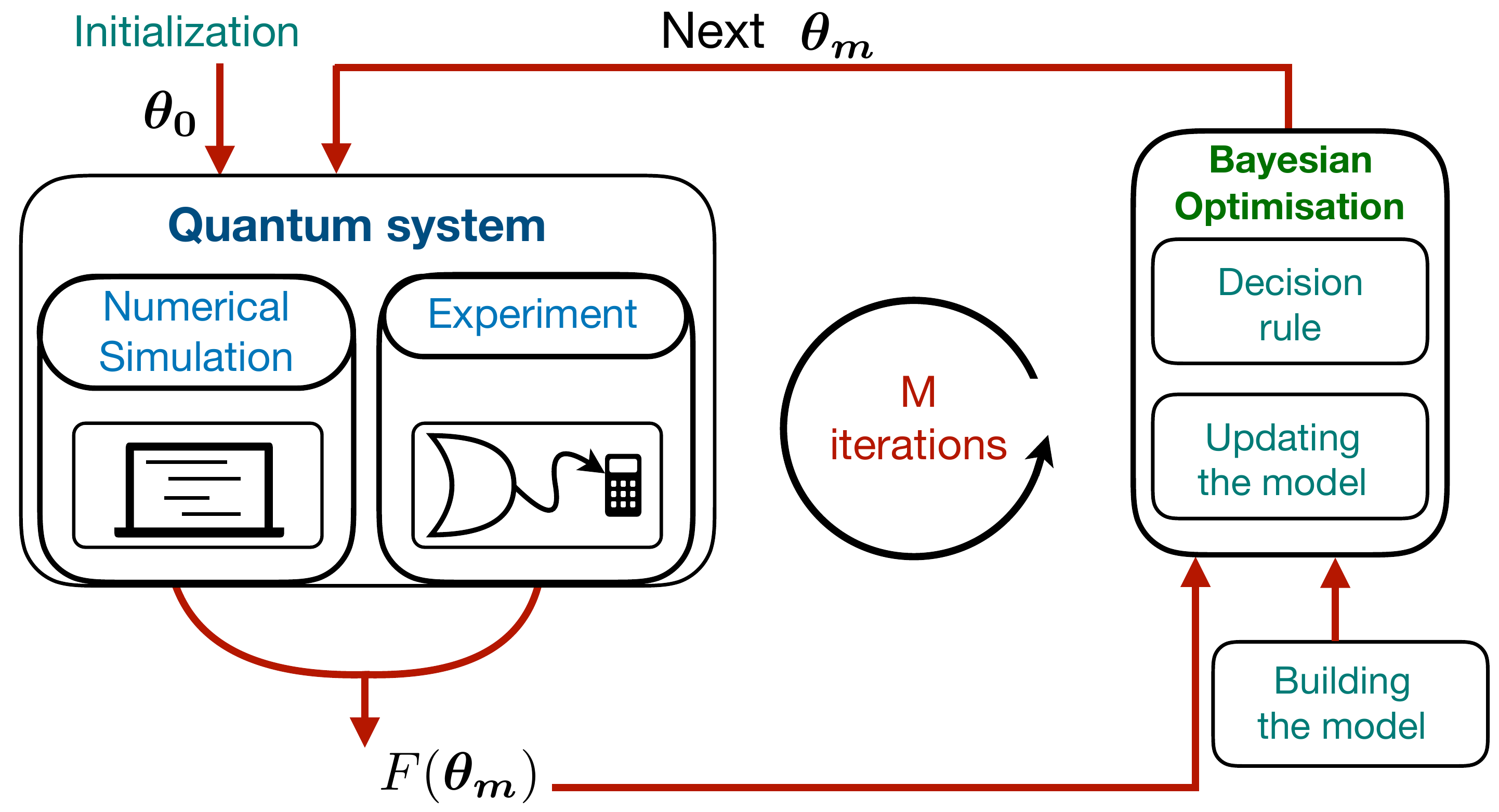}
\caption{Quantum optimal control scheme: $F(\boldtheta_m)$ is the FoM to maximize which depends on input parameters $\boldtheta_m$ that describes the control fields of the quantum system. $F(\boldtheta_m)$ is evaluated either numerically or experimentally. 
The optimization task is initiated by evaluating this FoM for a set of random control parameters $\boldtheta_0$. At every stage, the Bayesian optimizer suggests the next set of  parameters for the quantum system based on previous evaluations of the FoM. This process is repeated over M iterations.}
\label{QOCfig}
\end{figure}

\section{Quantum Optimal Control formalism}\label{CLQOLS}

Historically, quantum optimal control was first applied to manipulate chemical reactions by selectively breaking bonds in molecules using lasers \cite{rabitz, PhysRevLett.68.1500, assion1998control, QOC6, QOC7}; more recent examples involve quantum  metrology \cite{PhysRevLett.115.190801,poggiali2018optimal}, quantum computing \cite{kelly2014optimal,npj.45,li2017hybrid,feng2018gradient}, control over few qubits \cite{OCCJQ,PhysRevA.88.052315,PhysRevLett.99.170501,banchi2016quantum} as well as many-body systems \cite{KROTOV0a, BEC0, CRAB0}. 

Within the framework of optimal control, the entire optimization task involves maximizing the FoM,  $F(\boldtheta)$  with respect to the set of parameters $\boldtheta$ until an optimum solution, $\boldtheta^{\rm opt}$ is reached, as expressed below,
\begin{equation}\label{eq:QOC}
\boldtheta^{\rm opt} = \argmax_\boldtheta~F(\boldtheta) .
\end{equation}
$\boldtheta = [ \theta^1, \theta^2,\hdots,\theta^P]$ is a vector of size $P$ that represents the parametrization of the input control fields. For time-dependent control, this parametrization can be done using special functions \cite{GOAT}, Fourier series \cite{CRAB0} as well as piece-wise constant functions \cite{Khaneja2005296}. As illustrated in Fig.~\ref{QOCfig}, the FoM is evaluated for some random set of initial parameters $\boldtheta_0$. This is passed onto the optimizer which in turn suggests the next set of parameters, for which the FoM is again evaluated thus repeating the optimization task iteratively. At iteration $m$, $F(\boldtheta_m)$ is obtained either from a real experiment or through numerical calculations. The hope is to converge close enough to $\boldtheta^{\rm opt}$ well before exhausting the experimental/numerical resources. Although experimental constraints are not explicitly included in Eq.~(\ref{eq:QOC}), they can be incorporated either in the FoM directly or in the choice of the parametrization.

A variety of optimization algorithms are available and are broadly classified based on the ability to evaluate the gradient of $F(\boldtheta)$ with respect to $\boldtheta$. Gradient based algorithms such as the Krotov \cite{Krotov0, PhysRevLett.103.240501} and the \emph{gradient ascent pulse engineering} (GRAPE) \cite{Khaneja2005296} methods have been successfully applied to numerical simulations. Other gradient (or approximate gradient) based methods such as finite-differences \cite{Dive2018insituupgradeof}, \emph{simultaneous perturbation stochastic approximation} (SPSA) \cite{feng2018gradient, PhysRevA.91.052306} and an hybrid quantum-classical approach \cite{li2017hybrid, feng2018gradient} can be applied to experiments. These gradient based methods are considered as local optimization methods whose convergence is fast as long as the optimization landscape is well-behaved. However, these methods get compromised by the presence of any local minima and plateaus in the optimization landscape \cite{Zahedinejad,bukov2018RL, mcclean2018barren}. Alternatively, gradient-free algorithms are able to explore the optimization landscape more globally than gradient based methods making them less vulnerable to local minima. Among them, Nelder-Mead has been extensively used in the context of the \emph{chopped random basis} (CRAB) method \cite{CRAB0, CRAB1, DCRAB} as well as independently \cite{poggiali2018optimal,kelly2014optimal,egger2014adaptive}. It has the advantage of simplicity but can still get trapped in local minima and its convergence is limited by the presence of noise in the observations \cite{PhysRevA.91.052306}. Other popular non-gradient methods include evolutionary algorithms \cite{assion1998control,Zahedinejad,PALITTAPONGARNPIM2017116} and the recently introduced reinforcement learning techniques \cite{bukov2018RL, PhysRevX.8.031084, niu2019universal}. These methods usually require large number of iterations to find the optimal solution.

Bayesian Optimization (BO)  is a non-gradient based method offering an appealing alternative as it cleverly selects the next set of parameters to evaluate at each step of the optimization. This can lead to significant reduction in the number of iterations needed for convergence while performing global optimization. BO has the additional advantage of integrating probabilistic elements of data acquisition which can be exploited to further increase its efficiency \cite{sauvage2019}. These characteristics make it especially well suited to perform optimal control on quantum experiments but also when numerical simulations of the system are time-consuming.

\section{Bayesian Optimization}\label{BOSection}

The essential steps of BO used for quantum optimal control are shown in Fig.~\ref{QOCfig} and are discussed in this section: BO relies on an approximate model of the optimization landscape, which is updated at each iteration, and is leveraged to choose the next set of parameters for which the FoM is evaluated.

As highlighted in the previous section, the FoM can be obtained either experimentally or numerically and is passed to the optimization routine. Each of these evaluations can be time consuming, and in experimental setups, subject to noise. For these reasons, one cannot expect to resolve perfectly the true optimization landscape and probabilistic modeling becomes convenient. A suitable probabilistic model $f$ is specified in terms of a distribution, $p(f)$ taken to be a Gaussian process \cite{williams2006gaussian} which is detailed in \ref{prior}. In essence, $p(f)$  allows to favor smooth and regular functions to describe the unknown control landscape $F$. As this distribution does not yet incorporate evaluations of the FoM it is therefore referred as the \textit{prior distribution}.

The next step in BO is to update the probability distribution $p(f)$ based on the values of the FoM already collected. At an arbitrary step $D$ of the optimization, $D$ of such evaluation have been obtained and these set is denoted by a vector, $\mathbf{y}=[y_1(\boldtheta_1), \hdots, y_D(\boldtheta_D)]$. The \textit{posterior distribution} is the probability distribution $p(f|\mathbf{y})$ defined as the distribution for $f$ conditioned on all the evaluations $\mathbf{y}$ obtained so far and is specified using Bayes rule, 
\begin{equation} \label{eq:gpupdate}
p(f|\mathbf{y}) = \frac{p(f) p(\mathbf{y}|f)}{p(\mathbf{y})} ,
\end{equation}  
where $p(\mathbf{y})$ is the probability distribution for the set of observations $\mathbf{y}$ and $p(\mathbf{y}|f)$ is the likelihood for the set of observations $\mathbf{y}$ to occur based on a given model $f$. Specific details regarding the evaluation of the predictive distribution $p(f(\boldtheta)|\mathbf{y})$ for any control parameters $\boldtheta$ are described in \ref{posterior}.

Finally it remains to decide which set of control parameters $\boldtheta_{D+1}$ to use in the next step of the iterative optimization. One could naively choose it where the model $f$ takes its maximal mean value. However, this model is only approximative and thus likely to miss some interesting features of the optimization landscape, especially when the number of observations is low. Therefore it is also of interest to evaluate the FoM where uncertainty in the model is high. In BO, these two conflicting aspects, sometimes referred respectively as exploitation and exploration, are captured by an \textit{acquisition function} $\alpha(\boldtheta)$ and the next set of parameters is chosen where this function reaches its maximum. For example it could be taken as 
\begin{equation}\label{eq:bo:acq:ucb:main}
	\alpha(\boldtheta) = \mu_f(\boldtheta) + k \sigma_f(\boldtheta),
\end{equation}
with a positive scalar $k$ and where both the mean value $\mu_f(\boldtheta)$ and the standard deviation $\sigma_f(\boldtheta)$ are obtained according to the model predictive distribution given in Eq.~(\ref{eq:gpupdate}). The maximum of this function is reached when both its mean value and standard deviation, which quantifies the uncertainty in the modeling approach, are high (more or less emphasis on one or the other can be modulated with $k$). More details are given in \ref{decision}.

\begin{figure}
\centering
\includegraphics[width=1.0\textwidth]{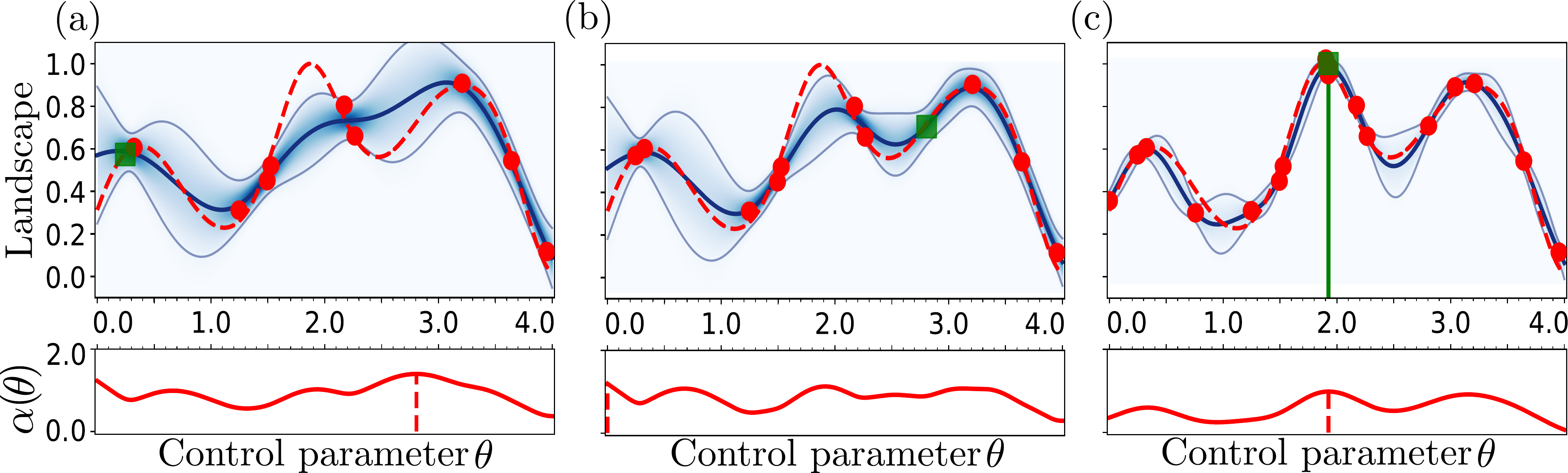}
\caption{Example of Bayesian optimization for a function of a single parameter $\theta$ after (a) $10$,
(b) $11$ and (c) $20$ iterations. The top panel depicts the underlying function $F$ to maximize (dashed red) along with its actual data points (red circles). The mean of the probabilistic model that approximates $F$ is represented by a solid blue line and a $95 \%$ confidence interval (shaded region delimited by the solid gray lines). The lower panel depicts the acquisition function, $\alpha(\theta)$ (red solid line) given by Eq.~(\ref{eq:bo:acq:ucb:main}) with $k=4$ (a-b) and 0 (c), whose maxima locations correspond to the selection of the next set of parameters in the optimization loop. For example, in (b) the new parameter was chosen at $\theta=2.7$ (shown in green box) which corresponds to the maximum of  $\alpha(\theta)$ in (a).}
\label{bofig}
\end{figure}
To illustrate the points discussed above, a simple example of one dimensional optimization is considered in Fig.~\ref{bofig}. The maximum value of an arbitrary function of a single parameter $\theta \in [0,4]$ is searched for using BO, implemented using the numerical package \cite{gpyopt2016}. This function represents the true FoM which is shown in dashed red and the noisy evaluations of the function, which serve as elements of $\mathbf{y}$, are shown as red dots. Ten of such evaluations have been obtained and the posterior distribution $p(f|\mathbf{y})$ is plotted in the top panel of Fig.~\ref{bofig}(a). The posterior distribution is defined by its mean (shown in solid blue line) and its variance  which are used to compute a $95\%$ confidence interval (depicted as shaded blue across the mean). The width of this interval results from the finite number of evaluations and the noisy evaluations.

As seen from Fig.~\ref{bofig}(a), on comparing the model with the true FoM, one finds that it does not adequately reproduce the underlying control landscape due to the small number of evaluations. This lack of knowledge of the true underlying landscape is captured by the variance of the distribution resulting in a large confidence interval. The acquisition function given in Eq.~(\ref{eq:bo:acq:ucb}) is plotted in the lower panel of Fig.~\ref{bofig}(a). In this case,  the maximum of the acquisition function is at $\theta \approx 2.7$ (highlighted as dashed vertical line), where the model has both, a high mean value (but not necessarily maximal) and high uncertainty. The FoM is then evaluated for this value of $\theta$. Fig.~\ref{bofig}(b) incorporates this new evaluation (green square) and exhibits the updated model based on a total of 11 iterations. This cycle of updating the model and suggesting the next parameter to evaluate is repeated another nine times and resulting in Fig.~\ref{bofig}(c). In this figure, the model is in close agreement with the true landscape and has successfully identified a parameter $\theta$ close to the true global maximum.

Before moving to the next section showcasing the results of BO applied to specific problems, we comment on the limitations and possible variations of this approach. Gaussian processes are a flexible tool to model well-behaved functions but are inadequate to model functions exhibiting discontinuities or varying degree of smoothness. However, this may be dealt with more complicated models as discussed in \cite{bernardo1998regression}. Similarly, approximating the true landscape well-enough may fail in this approach as could be the case with any model-based approach. The underlying noise model associated with the evaluation of the FoM, which enters in the likelihood term $p(f|\mathbf{y})$ in Eq.~(\ref{eq:gpupdate}) is assumed to be Gaussian in a standard BO implementation. There may be scenarios where this is not the case, nonetheless, it is possible to  incorporate non-Gaussian noise into the framework \cite{sauvage2019}. Another possible limitation is the computational complexity associated to evaluating the posterior distribution given in Eq.~(\ref{eq:gpupdate}) which grows as $D^3$ with $D$ as the number of accumulated evaluations. Alternatives to an exact treatment of Gaussian processes are discussed in the outlook which may help to alleviate this computational burden.

\section{Preparation of ordered states in ultra-cold systems beyond adiabatic methods}\label{orderorder}

Controlled preparation of many-body states with spatial correlations resembling solid-state matter is of great interest to the condensed matter community. For example, the SF-MI (Superfluid-Mott insulator) transition realized with bosonic atoms trapped in optical lattices \cite{SFTOMI0, SFTOMI1} simulates the transition of a conducting state to an insulator state in solid state physics. Similarly creating highly correlated ordered states with long range interactions in a cold gas of Rydberg atoms \cite{Pohl,van_Bijnen,Zeiher1} is akin to crystals in solids with long range Coulomb interactions between the electrons. In the Bose-Hubbard model, the only form of interaction is short-range while in the Rydberg system, it can be relevant over a range longer than a typical lattice spacing \cite{Bohlouli, Beguin}. Both of these examples are well studied in the context of ultra-cold physics and will serve as perfect testbeds for establishing the merits of BO techniques. 

In Sec.~\ref{MISF}, using the Bose-Hubbard model as a toy example, BO's ability to obtain an optimal protocol that drives the SF-MI transition is tested within the context of quantum speed limit \cite{PhysRevLett.103.240501}. Furthermore, BO's performance is benchmarked with other optimization methods in terms of its efficiency towards convergence. In Section \ref{RydbergCrystal}, BO is applied to the optimization of laser pulse dynamics to create Rydberg crystals with large fraction of Rydberg excitations in different lattice geometries; one-dimensional (1D), two-dimensional (2D) as well as three-dimensional (3D).

\begin{figure}
\centering	
\includegraphics[width=1.0\textwidth]{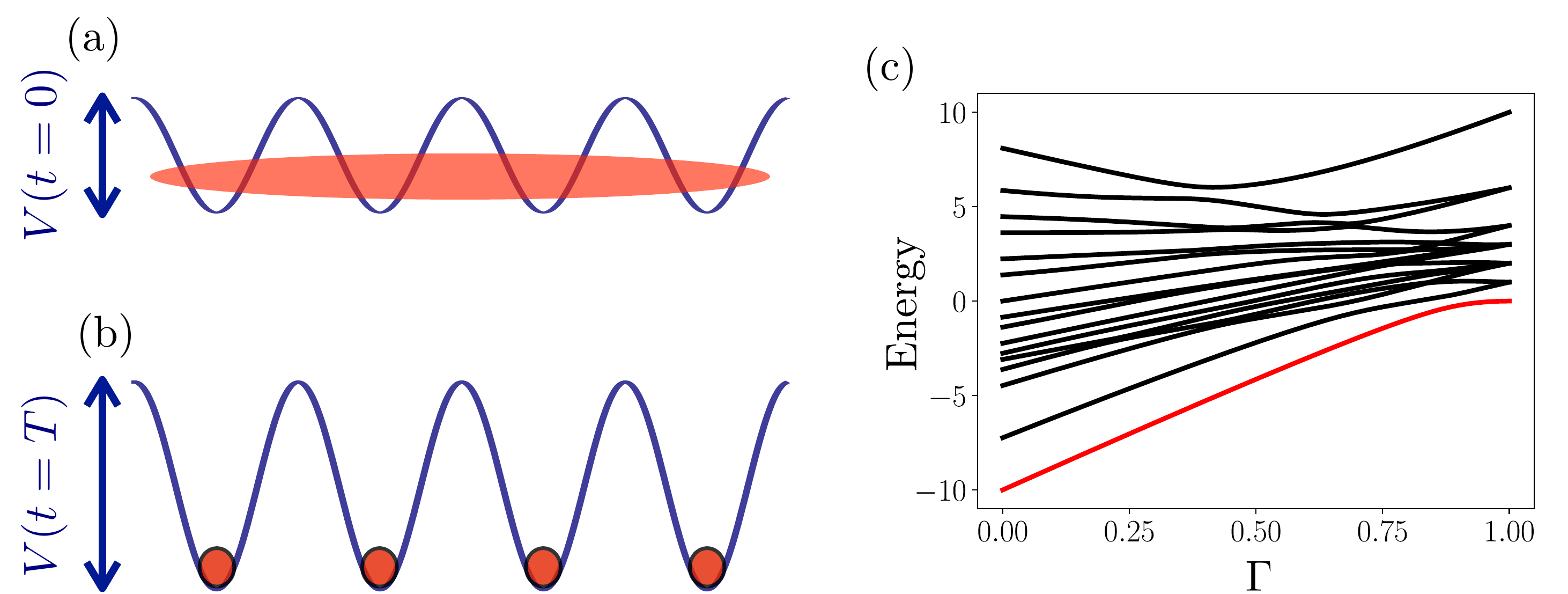}
\caption{Setup of Bose-Hubbard model: The goal of the optimization is to drive the system from an initial superfluid state illustrated in (a) to a Mott insulating phase as shown in (b) by dynamically changing the depth of the optical lattice $V(t)$. (c) Energy spectrum for the Hamiltonian in Eq.~(\ref{eq:res:bhm}) for different values of the control $\Gamma$.}\label{FigMISF}
\end{figure}
 
\subsection{Superfluid to Mott insulator transition in Bose-Hubbard model}\label{MISF}

The Bose-Hubbard model is widely studied in solid state physics and is simulated with bosonic atoms in an optical lattice. Although conceptually simple, this many-body system  cannot be mapped onto a single particle problem and contains the interesting phenomenon of transitioning from a SF state to a MI state \cite{SFTOMI0}. Experimental realization was first achieved \cite{SFTOMI1} by slowly varying the depth of the optical lattice potential. 

\subsubsection{Setup:}
For the Bose-Hubbard model, a homogeneous system of $N$ bosonic atoms with repulsive interactions in a lattice with $L$ sites is considered. The Hamiltonian for this model is given as 
\begin{equation}\label{eq:res:bhm_init}
\hat{\mathcal{H}}(t) = - J(t) \sum^L_{i=1} (\hat{b}_i\hat{b}_{i+1}^{\dagger} + \hat{b}_{i+1}\hat{b}_i^{\dagger}) + \frac{U(t)}{2} \sum^L_{j=1} \hat{n}_j(\hat{n}_j-1) .
\end{equation} 
The first term in the Hamiltonian describes the tunnelling of bosons between neighboring potential sites whose strength is given by $J(t)$. The annihilation (creation) of a boson at site $i$ is defined by operators $\hat{b}_i(\hat{b}_i^{\dagger})$ which follow the usual commutation relation, $[\hat{b}_i, \hat{b}_i^{\dagger}]  = \delta_{ij}$. The tunneling term tends to delocalize each atom over the lattice and the repulsive interaction between two bosonic atoms at each site is quantified by $U(t)$. Owing to the short range nature of the interactions compared to the lattice spacing, they are referred to as on-site interactions where  $\hat{n}_i=\hat{b}_i^{\dagger}\hat{b}_i$ is the number operator  for a given site $i$.  

As shown schematically in Fig.~\ref{FigMISF}(a-b), by varying the potential depth of the optical lattice in time,  both terms $J(t)$ and $U(t)$ are affected. $J(t)$ changes depending on the tunnelling barrier between neighboring lattice sites, while the change in the on-site interaction is due to the variation in atomic wave function confinement. Introducing a single dimensionless quantity, $\Gamma(t)= U(t)/(U(t)+J(t))$ such that $\Gamma(t)\in [0,1]$, the Hamiltonian is re-scaled as:
\begin{equation}\label{eq:res:bhm}
\hat{\mathcal{H}}(t) = - (1 - \Gamma(t)) \sum^L_{i=1} (\hat{b}_i\hat{b}_{i+1}^{\dagger} + \hat{b}_{i+1}\hat{b}_i^{\dagger}) + \frac{\Gamma(t)}{2} \sum^L_{j=1} \hat{n}_j(\hat{n}_j-1).
\end{equation}
The Bose-Hubbard Hamiltonian has two distinct ground states depending on the strength of the interactions $U$ relative to the tunneling  $J$. In the limit where $\Gamma(t)=0$, the interaction term vanishes and the ground state of each atom is delocalized over the entire lattice. There is uncertainty in the number of atoms per site and the many-body state can be described as a superposition of different atom number states, 
\begin{equation}\label{superfluid}
\ket{\psi_{SF}}\propto \Big( \sum_i^L b_i^{\dagger}\Big)^N|0\rangle . 
\end{equation}
Here $|0\rangle$ is the many-body vacuum state in the Fock basis representation. This SF state is characterized by a non-zero variance of the number operator and a constant global phase across the lattice.

In the opposite limit with $\Gamma(t)=1$, the hopping between adjacent sites is suppressed and the ground state of the system consists of localized atomic wave functions that minimize the interaction energy. For a homogeneous system, one can have commensurate filling of $n$ atoms per lattice site and the many-body state is then a product of local Fock states in the atom number for each lattice site. For the ground state where $n=N/L$, the Mott insulator is given as
\begin{equation}\label{mottinsulating}
\ket{\psi_{MI}}\propto \prod_{i}^L (b_i^{\dagger})^{N/L}|0\rangle .
\end{equation}
Global phase coherence of the matter wave field is lost at the expense of  perfect atom number correlations between lattice sites. In the next section we proceed to analyze the energy spectrum of the Bose-Hubbard model. Interestingly, in the limits of  $\Gamma(t)=0$ or $\Gamma(t)=1$, the Hamiltonian in  Eq.~(\ref{eq:res:bhm}) is completely integrable. However, for the intermediate values of $\Gamma$, the system is non-integrable \cite{Kolovsky_2004}.

\subsubsection{Many-body energy spectrum:}
For a chain of $L=5$ sites with periodic boundary conditions and unit filling ($L=N=5$) of bosons, the many-body energy spectrum as a function of $\Gamma$ is provided in Fig.~\ref{FigMISF}(c). This is solved using exact diagonalization in the Fock basis representation and is implemented with QuSpin \cite{quspin}. The energies $E_n$ of the many-body eigenstates $|n\rangle$ are plotted where $n=0,1,2\hdots$ enumerates each state in ascending order of energy at every value of $\Gamma$. The energies at $\Gamma=0$ and $\Gamma=1$ can be expressed analytically \cite{Kolovsky_2004}. The ground state is highlighted in red ranging from superfluid ground state ($\Gamma=0$) to Mott insulating ground state ($\Gamma=1$) and takes specific values, with energies $E_0(\Gamma=0) = -2N$ and $E_0(\Gamma=1) = 0$ respectively. There are multiple avoided crossings for all the other values of $\Gamma$. 

\begin{figure}
\centering	
\includegraphics[width=0.95\textwidth]{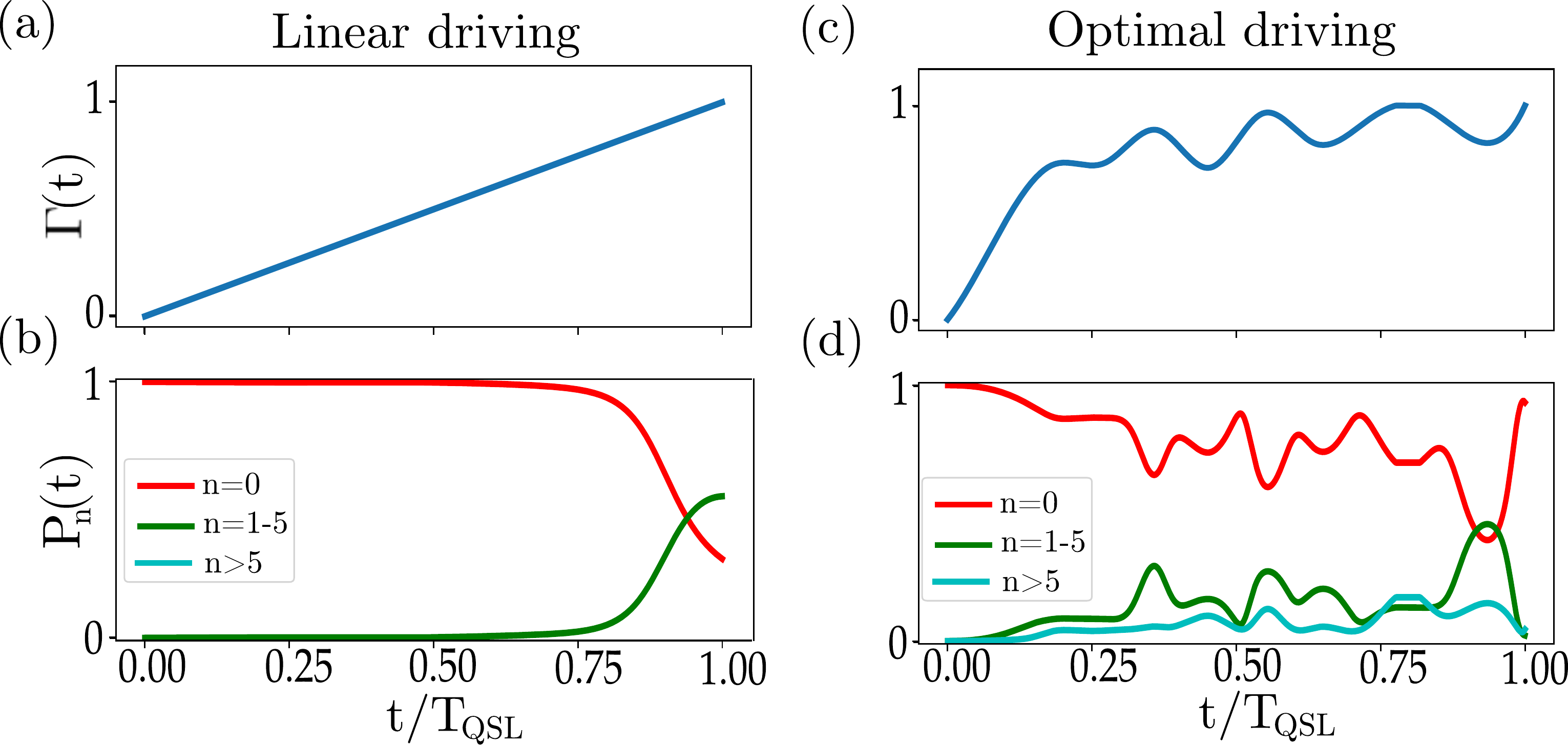}
\caption{Two controls are presented: a linear driving in (a) and an optimized one found by BO in (c). The evolution of the population $P_n(t)$ (defined in the main text) of the instantaneous eigenstates for the linear (b) and optimized protocol (d) are also reported. For clarity they are grouped by: ground state only (blue), first to fifth (green) and higher (turquoise) eigenstates.}\label{figsftomi_optim}
\end{figure}

\subsubsection{Optimization task:} The aim is to find an optimal control function $\Gamma^{opt}(t)$ that will efficiently drive an initial superfluid state as defined in Eq.~(\ref{superfluid})) to a Mott-insulating state (see Eq.~(\ref{mottinsulating})) in time $T$. The general set of parameters $\boldtheta$ described in Sec.~\ref{CLQOLS} take specific meaning in this context, as the time-dependent control $\Gamma(t)$ is parametrized with $10$ parameters representing the control values taken at equidistant time, $\Gamma(t = i \times T / 10)$ with $i \in \set{1,\hdots,10}$. With the boundary conditions, $\Gamma(0)=0$ and $\Gamma(T)=1$, the values of the control for intermediate times are obtained by fitting a cubic spline. For the FoM, we will first consider the fidelity to the target state. It is defined as $\mathcal{F}_{\rm MI} =|\langle \psi_{MI}\ket{\psi(T)}|^2$, where $\ket{\psi(T)}$ is the realized state at time $t=T$. While this fidelity is a natural choice for state-preparation task in numerical simulations, it is not readily accessible in an experiment. A second choice of FoM, $F_{exp}$, takes into account experimental constraints such as the effect of finite sampling. It is defined as $F_{\rm exp} = \langle V_i\rangle_{i}$, where $V_i$ is the variance in the occupation number estimated by averaging over for a given lattice site $i$.

In all cases the dynamics is solved numerically by exact diagonalization, and we denote the eigenstates of $\mathcal{H}(t)$ from Eq.~(\ref{eq:res:bhm}) as $|n(t)\rangle$. Thus, $|n(0)=0\rangle$ and  $|n(T)=0\rangle$ are the ground state SF and MI states respectively.

\subsubsection{Results and discussions:}

Using fidelity as the FoM, the duration of the protocol is fixed to the quantum speed limit as defined in \cite{van2016optimal,PhysRevLett.103.240501}, $T_{\rm QSL} = \pi/\Delta$, where $\Delta$ is the minimum energy gap between the ground and first excited state within the bounds of $\Gamma$ (see Fig.~\ref{FigMISF}(c)). This limit theoretically bounds the minimum time to reach perfect fidelity by the system. Fig.\ref{figsftomi_optim}(a) shows the drive with a simple linear increase of $\Gamma(t)$ and is compared to the optimal driving found with BO which is plotted in  Fig.\ref{figsftomi_optim}(c). Fig.\ref{figsftomi_optim} (b) and (d) exhibit the squared overlap of the state $|\psi(t)\rangle$ with $\ket{n(t)}$, defined as $P_n(t) = |\langle \psi(t)|n(t)\rangle|^2$ for the linear and optimized drives. As $\Gamma(t)$ is varied from 0 to 1, we retrieve the ground state fidelity at the end of the protocol with almost $90\%$ in the case of the optimized protocol while the linear ramp has a fidelity of less than $30\%$. For the linear protocol to reach the same fidelity as the optimized one, it would take about an order of magnitude longer in time thus establishing the non-adiabatic character of the optimum protocol.

\begin{figure}[t!]
\centering	
\includegraphics[width=1.0\textwidth]{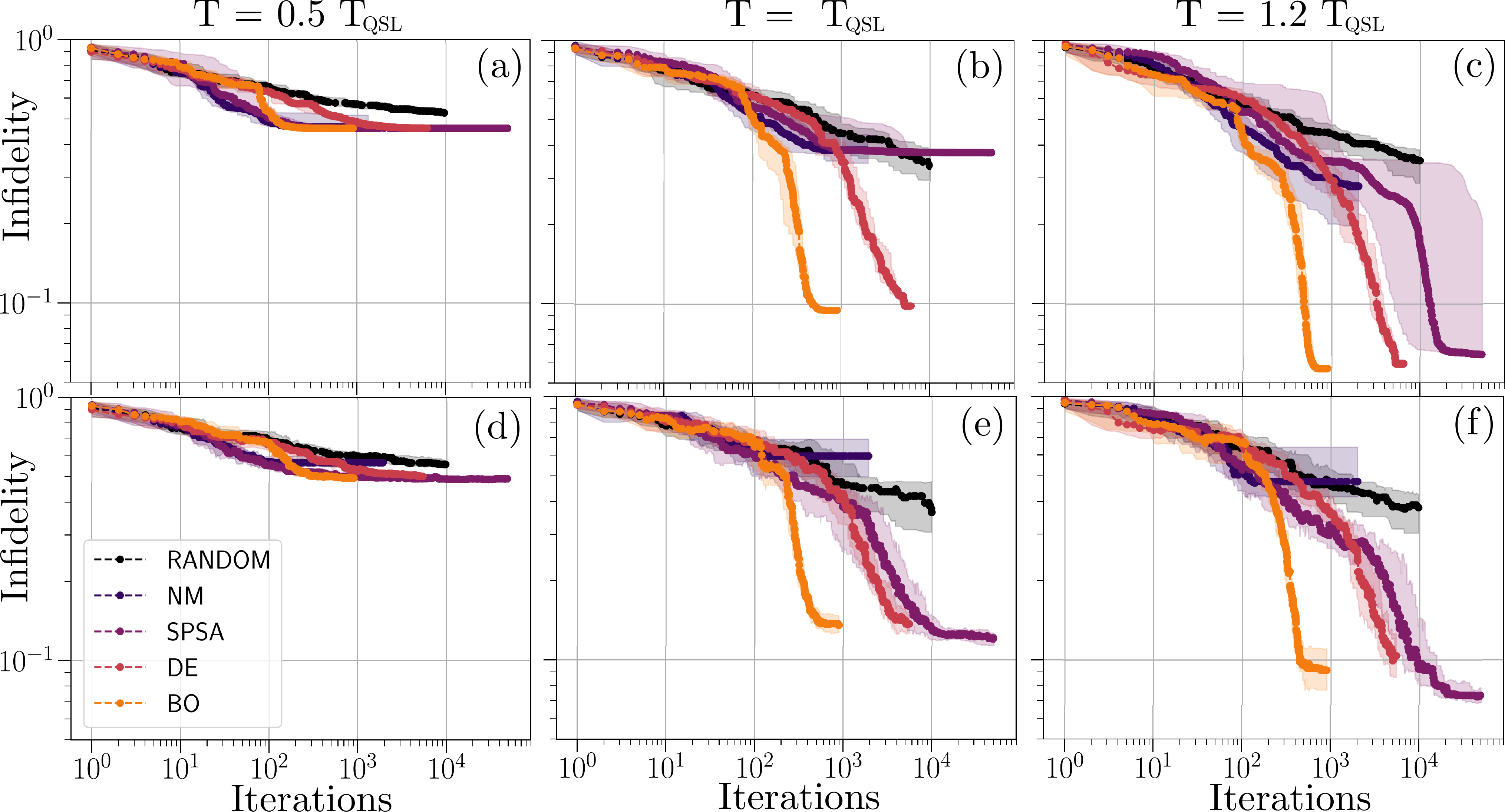}
\caption{The infidelity is reported as a function of the number of iterations for different optimization routines (colors in legend). All the results obtained are based on $30$ runs of each setup, for which the median (dots) and the first to third quartile interval (shaded region) are reported. Optimizations for different protocol times are shown: $T=0.5 \times T_{QSL}$(first column),  $T_{QSL}$ (second) and $1.2 \times T_{QSL}$(last), where $T_{QSL}$ is defined in the text. Two figures of merit are used: the optimizer has access to either exact evaluations of the fidelity (first row) or to a noisy (hence experimentally realistic) FoM (second row) as described in the main text.}	\label{figdiffmethods}
\end{figure}

To benchmark BO's overall performance for the SF-MI task, results obtained using BO are compared with other optimization routines (some of which were mentioned in Sec.~\ref{CLQOLS}) in Fig.\ref{figdiffmethods}. These include both gradient as well as non-gradient methods. For comparison with gradient based method, the spontaneous perturbation stochastic approximation (SPSA) algorithm was chosen \cite{SPSA} as it was shown to be competitive in the number of iterations for convergence and robust to noise \cite{PhysRevA.91.052306}. For comparison with non-gradient based methods, the differential evolution(DE) and Nelder-Mead(NM) optimizers were  implemented using Scipy \cite{scipy}. Finally random search (RANDOM) was also included in the list of optimizers. It relies on randomly sampling a new set of parameters at each iteration.  

In Fig.\ref{figdiffmethods}, the three columns correspond to optimization results for different protocol durations, $T=0.5T_{\rm QSL},~T_{\rm QSL},~1.2T_{\rm QSL}$, as these regimes can have different influences on the complexity of the optimization task \cite{Zahedinejad, bukov2018RL}. Fig.\ref{figdiffmethods}(a-c) show the infidelity, defined as $(1-\mathcal{F}_{\rm MI})$, as a function of the number of iterations. However for Fig.\ref{figdiffmethods}(d-f), a more experimentally realistic FoM is used, denoted by $F_{\rm exp}$. As mentioned before, this particular choice of FoM is motivated by experimental constraints such that the impossibility to project onto the target state and finite size sampling effects. The infidelity in this case is directly given by $F_{\rm exp}$ which is related to the variance in the occupation number at each lattice site. $F_{\rm exp}$ is minimized (vanishes) when a MI phase has been reached, thus providing an appropriate guidance for the optimization. Rather than the true expected value of the variance, an estimation of its value based on a finite number of repetitions (here 1000) is used.

As shown in Fig.\ref{figdiffmethods}, over all these different configurations, BO (orange curve) exhibits faster convergence towards low infidelity. Both SPSA and DE also converge to good parameters but at the expense of more iterations compared to BO. NM in most of the cases get quickly stuck in local minima with poor fidelities. When the time allowed for evolution is too short (see first column in Fig.~\ref{figdiffmethods}), the best final infidelity reached is of the order of only $50\%$. All optimizers are quite competitive in this case. However, when the time allowed for the transition to occur is more or equal to $T_{QSL}$ (second and third columns in Fig.\ref{figdiffmethods}), an order of magnitude less iterations are required by BO to converge to high fidelity control. Using the experimentally motivated FoM $F_{exp}$, the final fidelities dropped by $5\%$ for all the optimizers indicating additional difficulties to perform optimization in the noisy scenario. Interestingly SPSA performs better with the noisy FoM Fig.\ref{figdiffmethods}(d) while getting stuck when it has access to the true fidelity (c). This highlights the ability of SPSA to leverage noise to escape local minima. 

Based on these results, BO seems to provide a decisive advantage when it comes to number of iterations.

\subsection{Creation of Rydberg crystalline states}\label{RydbergCrystal}

After the first observations of interacting Rydberg gases \cite{Tong, Vogt, Reetz_Lamour_2008, Mohapatra, Raitzsch, Robert}, more control over the nature, form and structure of the quantum many-body state has been achieved by shaping the excitation sequence \cite{Schauss1455} and the arrangement of the atoms in specific patterns \cite{Labuhn}. Early proposals \cite{Pohl,van_Bijnen} for creating ordered phases in Rydberg gases were adiabatic in nature, in the sense that the overall dynamics was slower than the minimum energy gap of the many-body system.  Unfortunately, this energy gap decreases exponentially with the number of Rydberg excitations preventing the preparation of crystals with large number of Rydberg excitations.

Other proposals rely on admixing the ground state atoms with small amount of Rydberg state to create super-solid droplet crystals \cite{Henkel, Cinti}. This was experimentally implemented in \cite{Zeiher1}, but only for a very small system where the energy gap was rather large compared to the coherence time. Moreover Rydberg admixing of ground state atoms requires considerable fine tuning \cite{Balewski} especially to avoid spontaneous scattering and molecular resonances. More recently optimal control has been applied to optimize the preparation of Rydberg crystals \cite{Cui_2017} for specific geometries (quasi one-dimensional) and very low Rydberg fraction. Having proved BO's efficiency in the last example, there is a strong motivation to apply BO for preparing Rydberg ordered states by finding better protocols that give enhanced number of Rydberg excitations even in the presence of experimentally realistic parameter noise and possible lattice filling imperfections. 

\begin{figure}[t!]
\centering
\includegraphics[width=0.6\textwidth]{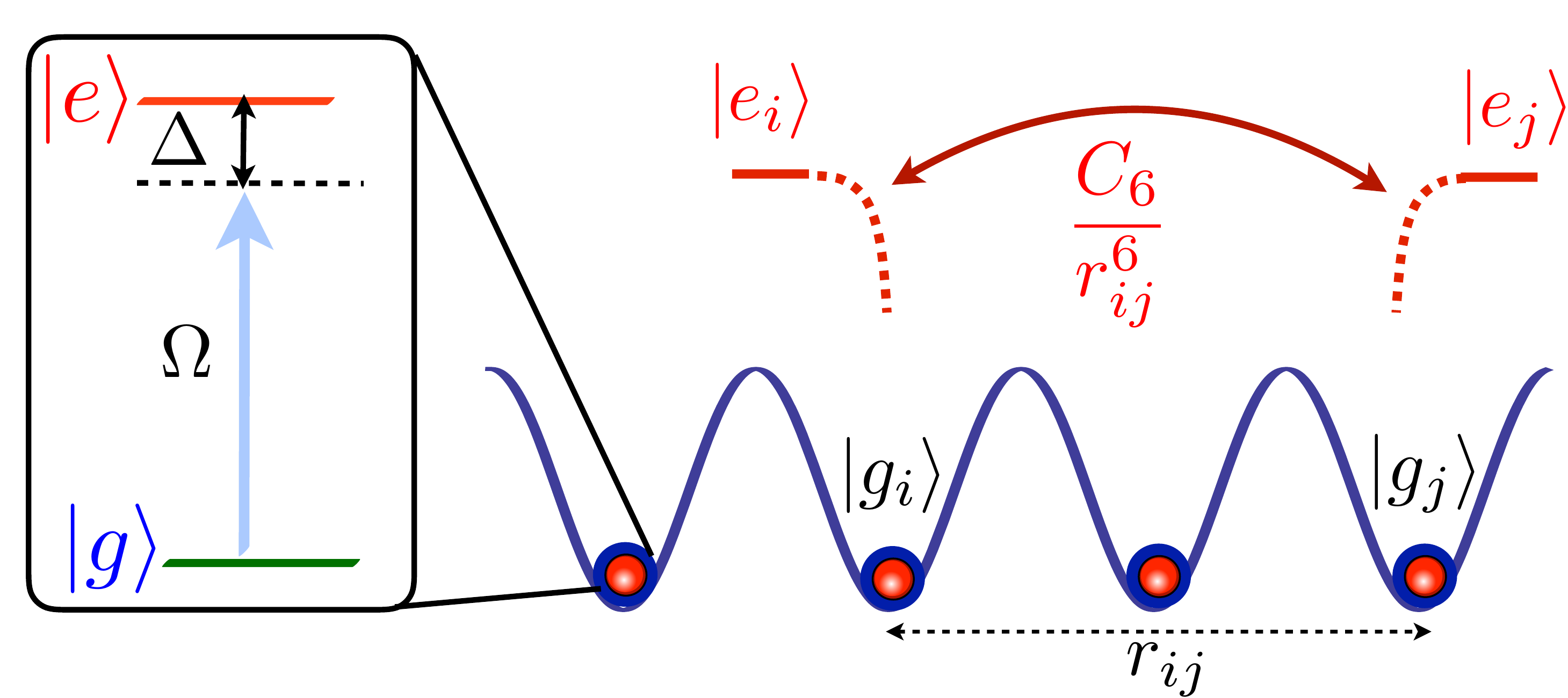}	
\caption{Schematic diagram of the setup: Ground state atoms denoted by $|g\rangle$ are trapped in a deep optical lattice. The ground state and Rydberg state ($|e\rangle$) of each atom is optically coupled with an effective Rabi frequency $\Omega$ and detuning $\Delta$. Two Rydberg atoms with a relative distance $r_{ij}$ interact via the van der Waals interaction.}
\label{FigRydSetup}
\end{figure}

\subsubsection{Setup:}\label{RydSetup} 
Fig.~\ref{FigRydSetup} illustrates the setup under consideration with Rb atoms in their ground state $\ket{g}=5S$ coupled to a particular Rydberg state $\ket{e}=50S$ with the help of a laser, thus treating each atom effectively as a  two-level system. The atoms can be either trapped in a deep optical lattice \cite{Schauss1455, Schauss} or in an array of optical dipole traps \cite{Nogrette} with uniform unit filling and lattice spacing $l=1.5$  $\mu$m. The advantage of trapping atoms at suitable spacings is that unwanted molecular resonances and scattering can be avoided. Moreover, if needed, simultaneous trapping of ground and Rydberg state atoms is possible in magic wavelength lattices \cite{Zhang,Topcu,Mukherjee_2011}. Two Rydberg atoms with positions $\mathbf{r_i}$ and $\mathbf{r_j}$ interact via the van der Waals interaction given by $V_{ij} = C_6(n)/|\mathbf{r_i}-\mathbf{r_j}|^6$, where $C_6>0$ is the van der Waals coefficient which  for our chosen Rydberg state with principal quantum number $n=50S$ takes the specific value, $C_6 = 1.56\times 10^{-26}$ Hz m$^6$  \cite{Reinhard}. The Hamiltonian describing the full setup in the rotating wave approximation is given as
\begin{equation}\label{Hatbasis}
\hat{H} = - \hbar\Delta \sum_i \ket{e_i}\bra{e_i} + \frac{\hbar\Omega}{2} \sum_i \left(\ket{g_i}\bra{e_i}  + {\rm h.c.}\right) + \sum_{i<j} V_{ij} \left(\ket{e_i}\bra{e_i} \otimes\ket{e_j}\bra{e_j} \right)
\end{equation}
where $\ket{e_i}\bra{e_i}$ is the projection operator onto the excited state for the i$^{\rm th}$ atom and $\ket{g_i}\bra{e_i}$ is the transition operator. The first two terms in the Hamiltonian represent the atom-laser interaction with the laser detuning $\Delta$ and the two-photon effective Rabi frequency $\Omega$ of the system. For small system sizes of less than ten atoms, a laser profile with uniform excitation is assumed such that all atoms experience the same Rabi coupling. There are different energy scales at play here where one can increase interaction strength $C_6$ by selecting higher principal quantum number $n$, which in turn provides more bandwidth for $\Omega$ and can be tuned with respect to the overall lifetime of the system. Any motional dynamics is neglected as their typical timescales are much longer than the fast excitation dynamics that is considered here. Spontaneous decay of the Rydberg state including black-body radiation for the chosen state, 50S is about $65$ $\mu$s \cite{Saffman_rev}. 

\begin{figure}[t!]
\centering
\includegraphics[width=0.9\textwidth]{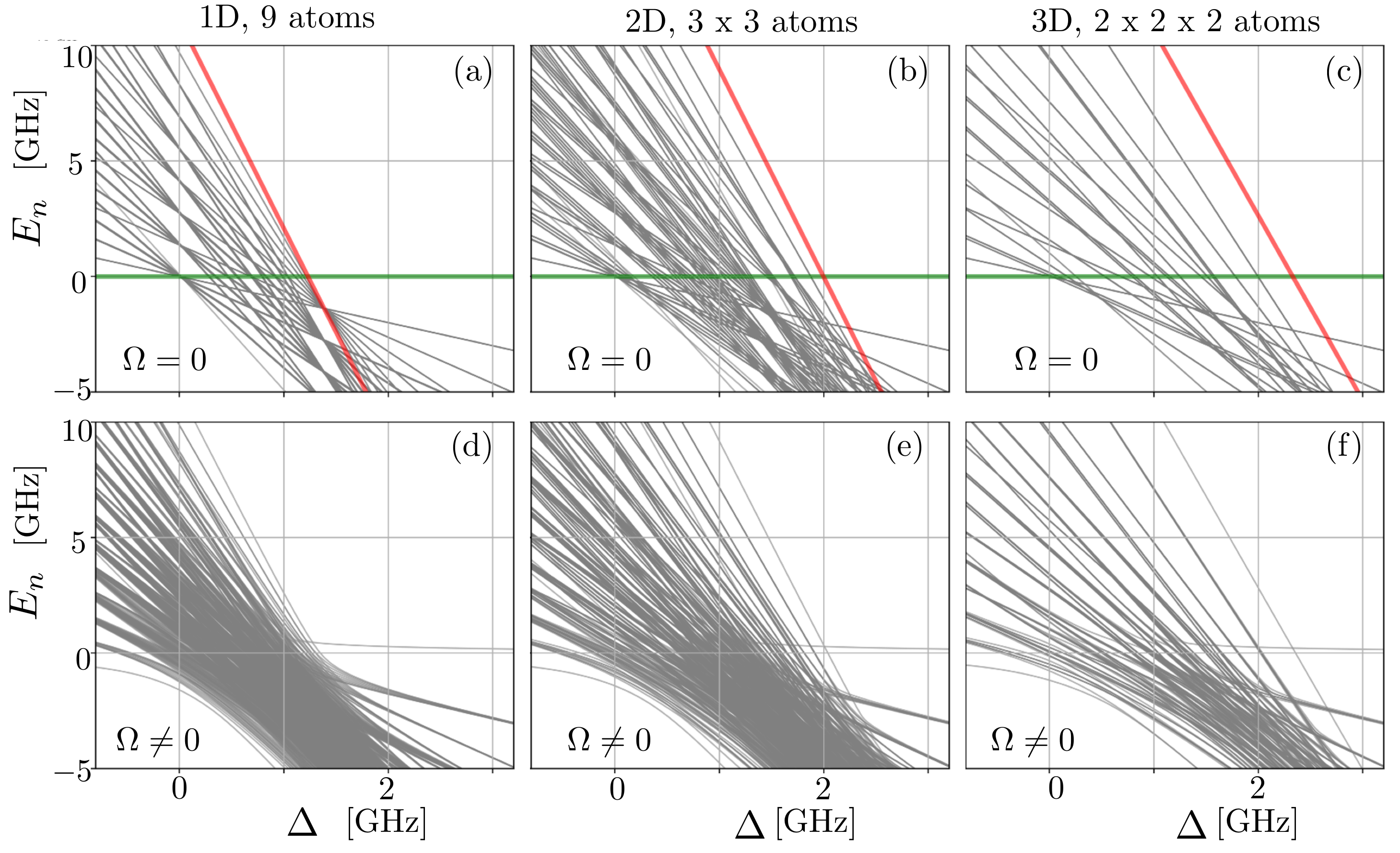}	
\caption{Panels (a-c)  correspond to zero laser intensity, many-body energy spectrum (given in Eq.~(\ref{mbeq})) with Rydberg state, $n=50$ (for which $C_6 = 1.56\times 10^{-26}$ Hz m$^6$) shown for different lattice geometries (with lattice spacing $l=1.5$  $\mu$m). The many-body ground and excited states are shown explicitly as green and red lines respectively. Panels (d-f) depict the many-body energy spectrum for the same parameters as (a-c) but now with a coupling of $\Omega=100$ MHz.}
\label{Figmanybody}
\end{figure}

\subsubsection{Many-body energy spectrum:}\label{manybody}

A gas of $N$ atoms with $n_e$ Rydberg excitations can be represented by the many-body eigenstate $|n_e,k\rangle$ where $k$ labels the different combinatorial possibilities to distribute $n_e$ excitations over $N$ atoms. The many-body energy spectrum with $\Omega=0$  for different geometries is shown in Fig.~\ref{Figmanybody}(a-c). In the zero intensity field scenario, the energy of a many-body state is given as
\begin{equation}\label{mbeq}
E_{n_e,k} = -n_e\Delta + \epsilon_{n_e,k}
\end{equation}
where $\epsilon_{n_e,k}$  are the interaction energies. The many-body ground state ($n_e=0$) has zero energy and is highlighted in green in Fig.~\ref{Figmanybody}(a-c).  States with the same number of Rydberg excitations form a manifold with the same slope with respect to detuning. Thus the state that has the highest slope (shown in red) corresponds to all atoms in the lattice excited, $n_e=N$. For non-zero laser intensity, anti-crossings occur between the many-body states as shown in Fig.~\ref{Figmanybody}(d-f). 

Since repulsive interactions are assumed for Rydberg atoms, the overall energy of the system is minimized by maximizing the spacing between Rydberg atoms and hence forming an ordered phase of Rydberg excitations. For a 1D system, sweeping the detuning from large negative values to specific positive values, the state with all atoms in the ground state gets connected to states with one, two or more Rydberg excitations depending on the final value of the detuning. One way to ensure that the final state is a Rydberg crystalline state, is to have the entire pulse dynamics performed slower than the inverse of the relevant energy gap. It is non-trivial to calculate the energy gap as it depends not only on the Rabi frequency but also on the interactions \cite{Pohl}. As the number of excitations increases, the energy gap decreases exponentially preventing the formation of Rydberg crystals with large Rydberg excitation fraction, $\nu = n_e/N \ll 1$. Moreover for 2D and 3D lattice geometries, the many-body energy spectrum is far more complicated as there are more degeneracies and requires further careful optimization over the $(\Omega,\Delta)$ control landscape. 

\subsubsection{Optimization task:} The dynamics for  Eq.~(\ref{Hatbasis}) is numerically solved using a linear multi-step integration method implemented using the QuTiP package \cite{JOHANSSON}. For this system, the relevant figures of merit are target state fidelity, spatial correlation function and number of Rydberg excitations, all of which can be potentially  measured in a typical Rydberg experiment with spatially resolved detection methods \cite{Ott_2016}. The control protocol is optimized for a specific number of  Rydberg excitations $n_e$ which is represented by the many-body eigenstates $\ket{n_e,k}$. By denoting the fidelity as $\mathcal{F}_{n_e}(t) =\sum_k|\langle\Psi(t)|n_e,k\rangle|^2$, the FoM provided at each iteration to BO is given by $\mathcal{F}_{n_e}(T)$. The sum is taken over all the configurations containing $n_e$ Rydberg excitations. 

The total duration of the dynamics is fixed to $T=1~\mu$s which is well below the overall lifetime of the many-body Rydberg system. The control field is parametrized using six parameters, $[\Omega (t_1), \Omega (t_2), \Omega (t_3), \Delta (t_1), \Delta (t_2), \Delta (t_3)]$ which correspond to Rabi frequencies and detunings at three different times. For intermediate times, the values of $(\Omega(t), \Delta(t))$ are interpolated using quadratic polynomial. Additionally, boundary conditions are imposed such that the intensity of the laser vanishes at $t=0$ and $t=T$. This is done by multiplying $\Omega(t)$ with a Tukey window function $w(t)$ \cite{Tukey}. The bounds for $\Omega\in[0,2.5~\text{GHz}]$ and $\Delta\in[-2.5~\text{GHz},4~\text{GHz}]$.

\subsubsection{Results and discussion:} 

\begin{figure}[t!]
	\centering
	\includegraphics[width=1.0\textwidth]{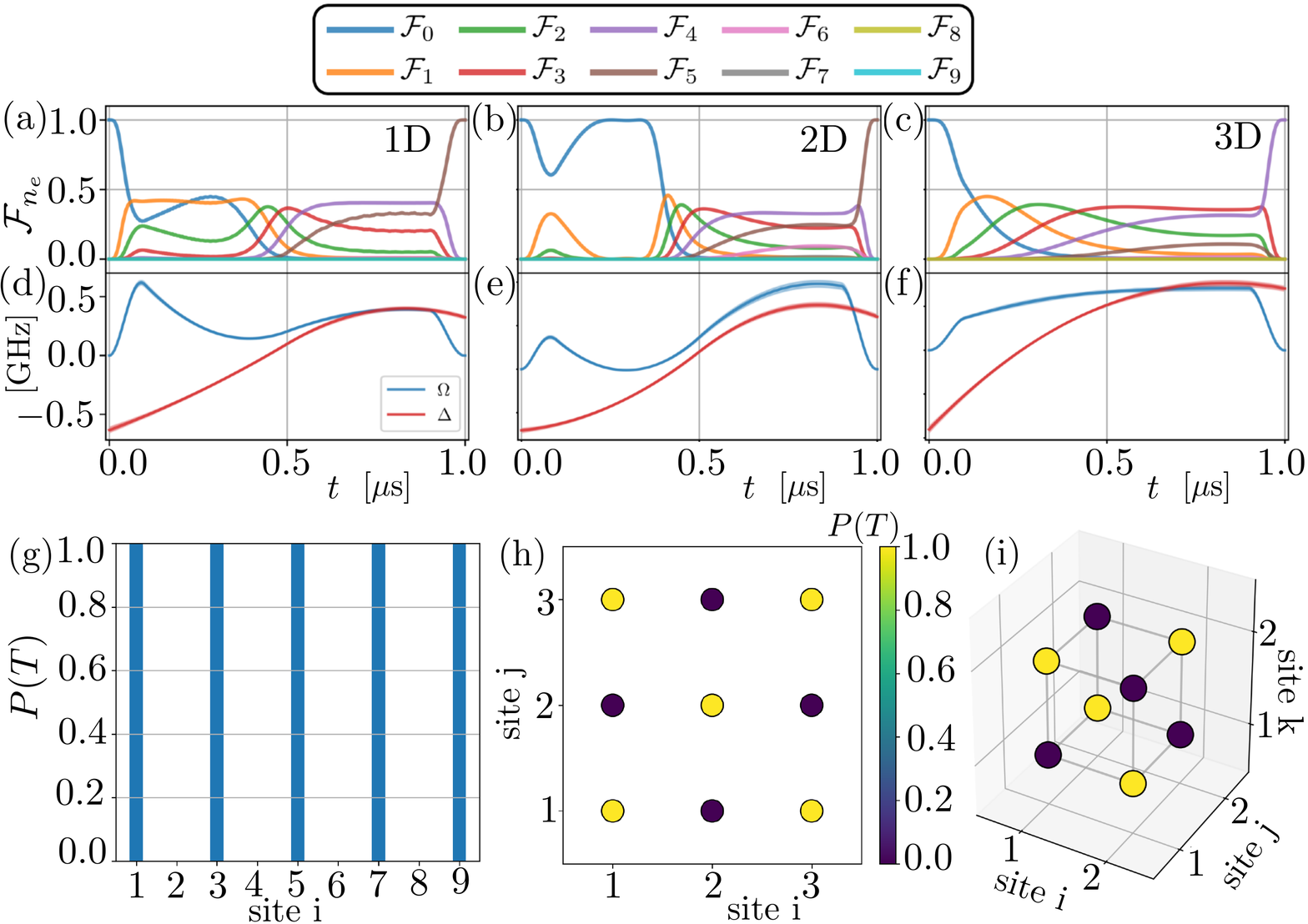}
	\caption{Results for optimized protocol dynamics using BO targeting a specific Rydberg excitations with different lattice geometries: (a-c) show the plots of fidelities $\mathcal{F}_{n_e}(t)$ (defined in the text) for a given number of Rydberg excitations $n_e$. It was optimized in (a) to maximize for $\mathcal{F}_5(t)$ in a 1D lattice, in (b) to maximize  for $\mathcal{F}_5(t)$ in a 2D lattice and in (c) to maximize  $\mathcal{F}_4(t)$ in a 3D lattice. Panels (d-f) are the corresponding optimized protocols given in terms of  $\Omega(t)$ and $\Delta(t))$. The three panels (g-i) show the final Rydberg excitation probability at any given lattice site corresponding to (a), (b) and (c) respectively.}
	\label{Rydprot}
\end{figure}

The total number of atoms for 1D and 2D lattices is nine and the protocol is optimized for five Rydberg excitations. The 3D lattice consists of eight atoms with the protocol optimized for four Rydberg excitations. The fidelities  $\mathcal{F}_{n_e}(t)$  over time for the optimal control protocols found by BO are depicted in Fig.~\ref{Rydprot} (a-c) and Fig.~\ref{Rydprot} (d-f) show the optimized control parameters for different lattice geometries. In all cases, BO achieves high fidelity for the selected number of excitations reaching convergence in the optimization within 10 iterations. In order to reproduce real experiments and test BO's resilience, 5\%  relative noise is introduced in $\Omega(t)$ and $\Delta(t)$ respectively. The standard deviation for $\Omega(t)$ and $\Delta(t)$ is shown in shaded blue and red colors respectively across their mean values (which are depicted with solid lines). It is most perceptible for $\Omega$ in Fig.~\ref{Rydprot} (e). These noises are usually negligible but could arise due to experimental drifts or Doppler broadening.

Inspecting the protocols optimized in Fig.~\ref{Rydprot}, it seems like BO naturally selects protocols with the general trend where the detuning starts from large negative values and then transitions to positive values. However, the temporal profiles of the optimized Rabi frequencies are non-trivial especially for the 1D and 2D lattice models. Fig.~\ref{Rydprot} (g-i) show the probability of Rydberg excitation $P(T)$ for every site at the end of the optimum protocol. It verifies the formation of Rydberg crystalline states as the atomic excitations are maximally separated for any given lattice model. For 1D and 2D lattices, there is only one unique spatial configuration that corresponds to ordered state with five Rydberg excitations. However in 3D lattice, there are two configurations that equally contribute to crystal with four Rydberg excitations. In Fig.~\ref{Rydprot} (i), one of the two configurations is rotated onto the other.

\begin{figure}[t!]
\centering	
\includegraphics[width=1.0\textwidth]{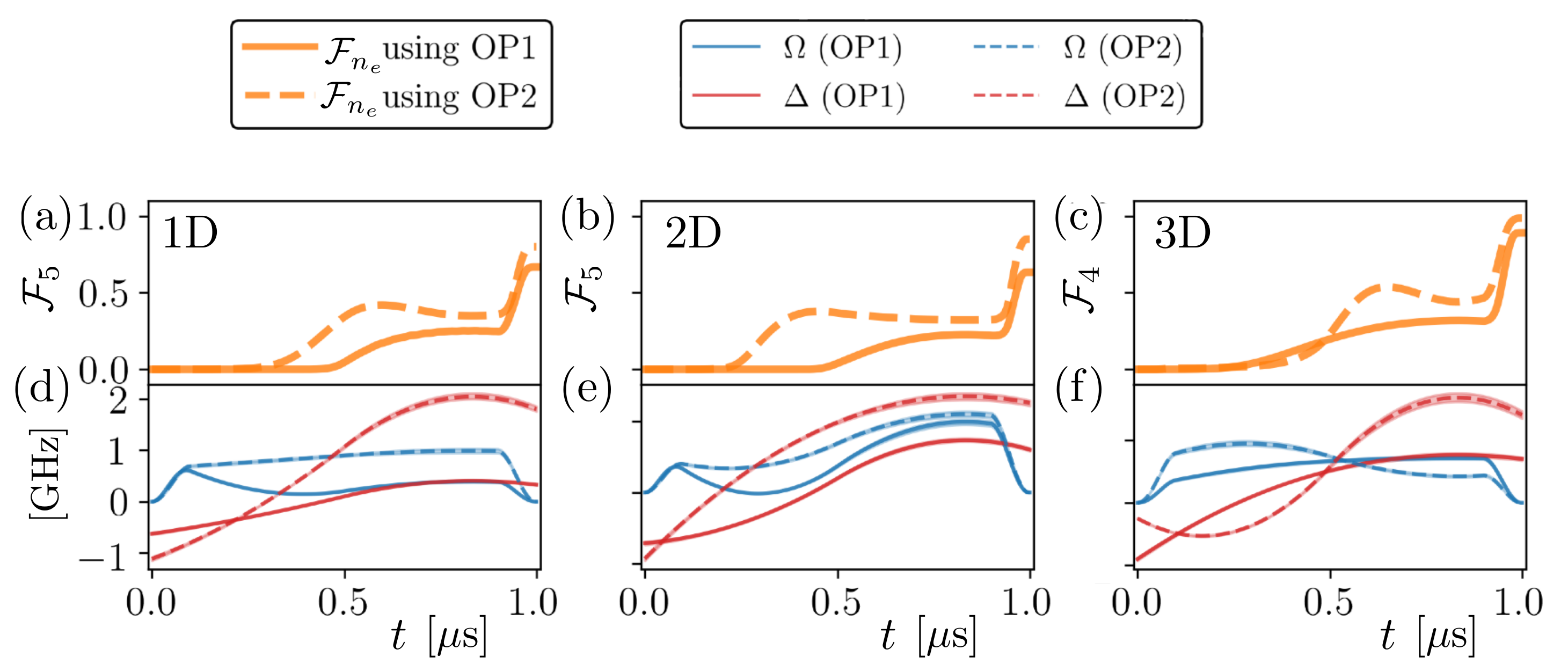}
\caption{Results of optimizations in the presence of imperfect lattice fillings for different lattice geometries. The solid (dashed) line refers to the protocol optimized for a perfect (imperfect) lattice filling which is denoted by OP1 (OP2). In (a-c), solid and dashed orange curves represent the fidelities for these specific protocols OP1 and OP2 both applied with imperfect lattice filling. (d-f) show the details of these protocols in terms of the Rabi frequency and detuning over time.}
\label{latticedefect}
\end{figure}

Errors originating from inaccurate readouts and imperfect lattice fillings are prevalent in lot of the current cold atom experiments \cite{Omran570, Schauss}. Both these imperfections are included in the simulations by associating a probability for Rydberg detection for each atom to be 0.9 when in Rydberg state. In one case, the optimum protocol is obtained for a perfect lattice and is labeled as OP1 while the other protocol is obtained for imperfect lattice filling which is labeled as OP2. The chosen FoM is the number of Rydberg excitations which was optimized for in Fig.~\ref{Rydprot}. 

Fig.~\ref{latticedefect} (a-c) show the FoM calculated by applying both control dynamics, OP1 (shown in solid orange) and OP2 (shown in dashed orange) on an imperfect lattice filling averaged over 50 realizations. The profiles of the drivings that account for lattice imperfections are vastly different from the ones obtained in the idealized case. Taking into account imperfections during the optimization results in better performances as can be seen in Fig.~\ref{latticedefect} (a-c) where OP2 exhibits higher final fidelities compared to OP1 across all lattice geometries. This shows the importance of performing optimizations directly onto the experiments rather than relying on idealized models or theoretical protocols for characterizing the noise sources and incorporating them in numerical simulations. Also the number of iterations required when noise is incorporated is five times higher which reiterates the need for optimization routines to converge fast in order to be of practical use. \\

BO found protocols to create Rydberg crystals in all three different lattice geometries with a Rydberg fraction of $\nu=n_e/N=0.55$ in $T=1~\mu$s. To create Rydberg crystals with such large Rydberg fraction would take much longer in an adiabatic scheme. For example, the adiabatic preparation of Rydberg crystals in one dimension were realized for  $\nu=0.016, T=1~\mu$s in \cite{Pohl} and $\nu=0.125, T=4~\mu$s in \cite{van_Bijnen}. The experiment \cite{Schauss1455} was done for $\nu=0.13, T=4~\mu$s  while the optimal control found using Nelder-Mead \cite{Cui_2017} was obtained for a quasi-1D system with $\nu=0.17, T=4~\mu$s. In summary, BO is able to efficiently obtain high fidelity protocols despite the parameter noise and lattice filling imperfections that were included in the simulations.

\section{Discussion and Outlook}\label{final}

Quantum optimal control aims at designing fast high fidelity schemes to prepare desirable target states in a quantum system. An optimization routine that is able to converge even when there is limited availability of data is advantageous for both, numerical simulations as well as for experiments. Along with this, the ability of the optimization routine to handle noisy data is highly beneficial for experimental setups. In this context, BO seems to be a promising optimization tool and to make this case, it was successfully applied to the creation of spatially ordered phases of ultra-cold gases trapped in lattices.

In the SF-MI transition task, the performance of BO in comparison to other optimization routines was found to be the most efficient with regards to the number of iterations needed for convergence. This is primarily due to its ability to cleverly select the next set of parameters to probe. In the second example, notwithstanding the complexity in the many-body energy spectrum and the vanishing energy gap for large Rydberg systems, BO obtained control protocols to create Rydberg crystalline states in lattice models of arbitrary dimension with much larger Rydberg fraction than previous methods. More interestingly, BO successfully navigated the optimization landscape even with the inclusion of noisy parameters and imperfect lattice fillings which is evidence of the advantage in accounting for noisy data in the modeling approach.

In the examples studied here, the parametrization of the control field was taken to be either spline or quadratic functions. Other types of smooth parametrizations, for example with Fourier components or even with Gaussian processes, could also be tested. Furthermore in all the optimizations performed in this work no initial guess function was assumed for the control. Such a guess could be directly taken into account in the parametrization of the control function \cite{CRAB1} and may result in optimized controls with higher fidelity. Similarly the number of parameters used for the time-varying control field could be increased to allow for finer control. However, in practice optimizing over higher dimensional parameter spaces may be challenging. For a given number of iterations, a trade-off between flexibility in the control field and realistic number of iterations has to be found.

Alternatively it would be interesting to study iterative refinements of the control field keeping the number of parameters fixed and small at each stage. For example, during the optimization, it may be realized that some temporal parts of the control field are more important than others and a new parametrization could be tested accordingly. Another option would be to repeatedly optimize over additive perturbations of the control function, each with a randomized parametrization, as is done in \cite{DCRAB}. Such iterative procedures could be made experimentally feasible if used in conjunction with the low data requirement provided by BO. 

Finally, we would like to remark that the field of probabilistic machine learning \cite{Ghahramani2015} is advancing at a fast pace and has a lot to offer for the characterization and optimization of quantum systems. In particular several alternatives for the model used here for BO have been suggested \cite{hensman2013gaussian, snoek2015scalable}. They could be easily integrated into the framework and decrease the computational burden associated with Gaussian processes, especially when the number of iterations becomes really large. Another interesting direction of research for even more efficient optimizations could originate from the field of transfer-learning \cite{swersky2013multi}. For example, it could allow to leverage results of optimizations performed on an (approximative) numerical simulator to speed up optimizations performed on the real experiments.

\section{Acknowledgments}
The authors acknowledge H. Weimer for insightful and productive discussions. Both Imperial and Stuttgart have received funding from the QuantERA ERANET Cofund in Quantum Technologies implemented within the European Union’s Horizon 2020 Programme under the project Theory-Blind Quantum Control TheBlinQC and from EPSRC under the grant EP/R044082/1. This work was supported through a studentship in the Quantum Systems Engineering Skills and Training Hub at Imperial College London funded by the EPSRC(EP/P510257/1).\\

\bibliographystyle{iopart-num}

\bibliography{references}

\appendix	
\section{Bayesian optimization: techniques}\label{sec:appendix:bo}

The problem of quantum optimal control defined in Section.\ref{CLQOLS} consists on finding optimal (or good enough) control parameters $\boldtheta$ such that the figure of merit $F(\boldtheta)$, is maximized. BO relies on an approximative model $f$, called the \textit{surrogate model}, of the figure of merit which is used to guide the optimization. 

\subsection{Building the surrogate model: Prior distribution}\label{prior}
 
 In the context of BO the surrogate model is taken to be a \textit{random function}, also known as \textit{stochastic process}.
Random functions extend the notion of a finite set of random variables to an infinite one. Thus, considering $f$ to be a random function means that for any of the infinitely many input parameters $\boldtheta_m$, the value $f(\boldtheta_m)$ taken by the function is itself a random variable. 
More precisely, $f$ is taken to be a \emph{Gaussian Process} (GP) which is a specific type of random functions \cite{williams2006gaussian}. A GP is the natural extension of a Multi-Variate (MV) Gaussian distribution: while a MV Gaussian distribution is entirely specified by a \emph{mean vector} and a \emph{covariance matrix}, a GP is given in terms of a \emph{mean function} $m$ and a \emph{covariance function} $k$. These two functions define respectively the mean $ \langle f(\boldtheta_m) \rangle = m(\boldtheta_m)$, and the covariances $\langle f(\boldtheta_m) f(\boldtheta_n) \rangle - \langle f(\boldtheta_m) \rangle \langle f(\boldtheta_n) \rangle = k(\boldtheta_m, \boldtheta_n)$ of the values taken by the model at any input parameters $\boldtheta_m$ and $\boldtheta_n$. Considering the model $f$ to be a GP with mean function $m$ and covariance function $k$ is denoted as $p(f)=\mathcal{G}\mathcal{P}(m, k)$ and is defined such that \emph{any} finite collection of random variables $f(\boldtheta_i)$, of arbitrary length $N$, follows a multivariate Gaussian distribution:
\begin{equation}\label{eq:bo:gpdef}
	p \Big(\begin{bmatrix}
		f(\boldtheta_1) \\
		\vdots\\
		f(\boldtheta_N)
	\end{bmatrix} 
	\Big)
	= \mathcal{N} \Big(
	\mathbf{m} =
	\begin{bmatrix}
		m(\boldtheta_1) \\
		\vdots\\
		m(\boldtheta_N)
	\end{bmatrix}
	,  
	K = 
	\begin{bmatrix}
		k(\boldtheta_1, \boldtheta_1) &..& (\boldtheta_1, \boldtheta_N) \\
		\vdots& & \vdots\\
		k(\boldtheta_N, \boldtheta_1)& ..&(\boldtheta_N, \boldtheta_N)
	\end{bmatrix} \Big)
\end{equation}
where $\mathcal{N}$ denotes a Gaussian distribution here with a mean vector $\mathbf{m}$ of length $N$ and a covariance matrix $K$ of dimension $N \times N$. 
The distribution over functions $p(f)$ or equivalently the joint distributions over finite sets of function values as given in Eq. \ref{eq:bo:gpdef}, are referred as the prior distribution as they do not incorporate any observations yet. 

\begin{figure}
\centering
\includegraphics[width=0.95\textwidth]{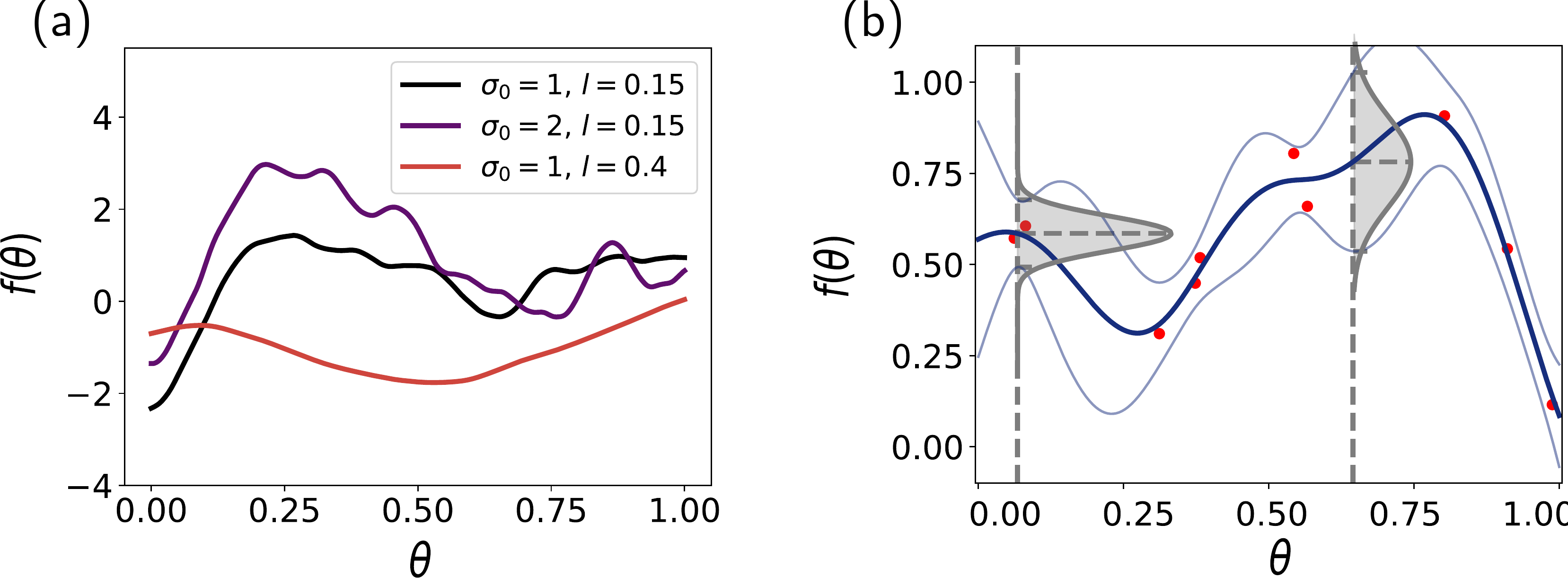}
\caption{(a) The figure shows three functions sampled from a GP with a Mat\'{e}rn covariance (Eq.\ref{eq:bo:matern52}) for different values of the length-scale $l$ and the variance parameter $\sigma_0^2$. (b) Predictive distribution obtained according to Eq.\ref{eq:bo:predictive} for a 1-d parameter $\theta \in [0,1]$ with $10$ noisy observations (red dots). The distribution is displayed in terms of its mean function $\mu_f(\theta)$ (blue thick line) and a $95\%$ confidence interval (grey light envelope) with boundaries $\mu_f(\theta) \pm 1.96 \times \sigma_f(\theta)$. Additionally examples of the full density for two specific inputs $\theta=0.25$ and $\theta=2.6$ are explicitly shown (vertical grey shaded slices).}
\label{hp}
\end{figure}
A specific choice of the mean and covariance functions defines the global properties of the model. The mean is often taken to be vanishing, $m(\boldtheta) \equiv 0$ and the prior assumptions on the function $f$ are entirely delegated to the choice of the covariance function $k$. A common choice, followed here, for the covariance function is the \emph{Mat\'{e}rn 5/2} function imposing the model to be twice differentiable \cite{williams2006gaussian}. This \emph{Mat\'{e}rn 5/2} function is defined as
\begin{align}\label{eq:bo:matern52}
k_{5/2}(\boldtheta_m, \boldtheta_n) = k_{5/2}(x = |\boldtheta_m - \boldtheta_n|) = \sigma_0^2(1+\frac{x}{l}+\frac{x^2}{3l^2}) e^{-x/l},
\end{align}
where the variance $\sigma_0^2$ and the length-scale $l$, are free hyper-parameters. Essentially the variance $\sigma_0^2 = k_{5/2}(x=0)= \langle f^2(\boldtheta_m) \rangle$ specifies to which extend any variable $f(\boldtheta_m)$ is expected to deviate from its mean value (here $0$ with our choice of mean function $m$), while the length-scale $l$, appearing in the exponentially decaying term, scales the distance $x$ between the parameters $\boldtheta_m$ and $\boldtheta_n$ which is indicative of the extent of the correlations between different values of the function. To illustrate the impact of these hyper-parameters, Fig.\ref{hp} (a) shows $3$ functions with have been sampled with different hyper-parameters values (given in legend). Each sample is obtained according to Eq.~(\ref{eq:bo:gpdef}) for a vector of $1$-d input parameters $[\theta_1, \hdots, \theta_N]$ with $N$ large enough such that each sample with finite size N effectively looks like a function. 

In general, it is not possible to have a precise idea of the values of these hyper-parameters beforehand but they can be fitted to the observations at any stage of the optimization. This fitting is often done by minimizing the log marginal-likelihood \cite{williams2006gaussian} and is implemented in any GP library such as \cite{gpy2014}. In summary the prior distribution is entirely defined by the choice of a GP, a mean function $m$, a covariance function $k$ and the hyper-parameters $\sigma_0^2$ and $l$. In particular, under the choice of a mean and a kernel function presented here, a priori each value $f(\boldtheta_m)$ taken by the model has a distribution: :
\begin{equation}\label{eq:prior}
p(f(\boldtheta_m))=\mathcal{N}(0, \sigma_0^2).
\end{equation}

	


\subsection{Updating the surrogate model: Predictive distribution}\label{posterior}
Starting with the prior distribution described in the previous section, it remains to incorporate observations to the model to obtain the posterior distribution. In the present context this posterior is also called the predictive distribution as it allows to make prediction for unseen parameters. Observations are denoted by a vector $\mathbf{y} = [y(\boldtheta_1),\hdots,y(\boldtheta_M)]$ where each element $y(\boldtheta_m)$ corresponds to an evaluation of the figure of merit at iteration $m$ for parameters $\boldtheta_m$, over a total of $M$ iterations.

In the case of noiseless evaluations of the figure of merit one directly records the values taken by the model for the different parameters tried. Writing this vector of values $\mathbf{f} = [f(\boldtheta_1), \hdots, f(\boldtheta_M)]$, it follows the equality $\mathbf{f}=\mathbf{y}$. Applying the definition of a GP in Eq.~(\ref{eq:bo:gpdef}) to the vector of random variables $[f(\boldtheta_1),..., f(\boldtheta_M), f(\boldtheta^*)]$ and using the conditioning properties of Gaussian distributions \cite{williams2006gaussian}, the predictive distribution is given by
\begin{equation}\label{eq:bo:predperfect}
	p(f(\boldtheta^*) | \mathbf{f}) = \mathcal{N}(\mu_f(\boldtheta^*) = \mathbf{k}^T K^{-1} \mathbf{f}, \sigma^2_f(\boldtheta^*) = \sigma_0^2 - \mathbf{k}^TK^{-1}\mathbf{k}),
\end{equation}
where the column vector $\mathbf{k}$ has entries $\mathbf{k}_m=k(\boldtheta_m, \boldtheta^*)$, and elements of the covariance matrix $K$ are given by $K_{m,n}=k(\boldtheta_m, \boldtheta_n)$. The symbols $\mu_f(\boldtheta^*)$ and $\sigma_f(\boldtheta^*)$ denote the mean and standard deviation of this predictive distribution. They both are function of the parameter $\boldtheta^*$ as each element of $\mathbf{k}$ depends on it. Compared to the prior distribution $p(f(\boldtheta^*))$ in Eq.~(\ref{eq:prior}), the mean of the predictive distribution has been shifted from $0$ to $\mathbf{k}^T K^{-1} \mathbf{f}$ and its variance has decreased from $\sigma^2_0$ by a positive quantity $\mathbf{k}^TK^{-1}\mathbf{k}$, resulting from the incorporation of the observations. The most computational demanding part of evaluating this mean and variance comes from the inversion of the matrix $K$ which has complexity $\mathcal{O}(M^3)$.

In an experimental scenario, the set of measurements $\mathbf{y}$ does not directly reveal the values taken by the model, still model and observations can be related by positing a noise model. This noise, originating from both experimental imperfections and also quantum fluctuations, can be approximated as an additive constant Gaussian noise such that $y(\boldtheta_j) = f(\boldtheta_j) + \varepsilon_j$, where the noise terms $\varepsilon_j$ are assumed to be independently drawn from the same distribution $p(\varepsilon_j) = \mathcal{N}(0, \sigma_N^2)$. The extra parameter $\sigma_N$, capturing the amount of noise, can be fixed or could also be fitted to the data in the same way as the hyper-parameters $\sigma_0^2$ and $l$. Under the independence assumption for the noise terms, the likelihood of recording the full data set of observations $\mathbf{y}$ for a given set of values $\mathbf{f}$ taken by the model can be written as $p(\mathbf{y}|\mathbf{f}) = \mathcal{N}(\mathbf{f}, \sigma^2_N I)$. With this simple phenomenological model of the noise the sought-after predictive distribution can be derived:
\begin{align}
\begin{split}
\label{eq:bo:predictive}
p(f(\boldtheta^*)| \mathbf{y}) &= \int \mathbf{df} p(f(\boldtheta^*) | \mathbf{f}) p(\mathbf{f}|\mathbf{y}) = \int \mathbf{df} p(f(\boldtheta^*) | \mathbf{f}) p(\mathbf{y}|\mathbf{f}) p(\mathbf{f})/p(\mathbf{y})\\
&=\mathcal{N} \Big(\mu_f(\boldtheta^*) = \mathbf{k}^T (K+\sigma^2_N I)^{-1} \mathbf{f},\; \sigma^2_f(\boldtheta^*) = \sigma_0^2 - \mathbf{k}^T(K+\sigma^2_N I)^{-1}\mathbf{k} \Big),
\end{split}
\end{align}
where the first line shows how it can be decomposed, using Bayes rule and marginalization over the vector of values $\mathbf{f}$, as a product of the prior distribution $p(\mathbf{f})$ appearing in Eq. \ref{eq:bo:gpdef}, the conditional distribution in Eq.~(\ref{eq:bo:predperfect}), the likelihood $p(\mathbf{y}|\mathbf{f})$ previously defined and the probability of obtaining a given set of observations $p(\mathbf{y})$. The next line provides the analytical result of this integral \cite{williams2006gaussian}. This final result is quite similar to Eq.~(\ref{eq:bo:predperfect}) except that the inverse of the covariance matrix $K^{-1}$ has now been replaced by $(K+\sigma^2_N I)^{-1}$ thus incorporating noise in the observations. 

A practical example of evaluating this predictive distribution for a one-dimensional parameter $\theta \in [0,4]$ based on a set of $M=10$ observations is provided in Fig.~\ref{hp}(b). The mean function $\mu_f$ and a confidence interval with bounds $\mu_f \pm 1.96 \sigma_f$ have been represented with, in addition, the explicit densities $p(f(\theta)| \mathbf{y})$ for two specific values of the parameter. Far away from the observations, for example for $\theta\approx0.75$, the confidence interval is wider indicating uncertainty in the predictive distribution, while closer to observations, for example for $\theta\approx0.5$, it shrinks but does not vanish as the model identified noise in the observations (more precisely the value of the hyper-parameter $\sigma_N$ representing the strength of noise after fitting is non $0$).

\subsection{Decision rule: Acquisition function}\label{decision}
The last step in the BO framework is the choice of the next parameter $\boldtheta_{M+1}$ to observe based on the surrogate model. This new parameter could be picked such that it maximizes the mean of the predictive distribution given in Eq.~\eqref{eq:bo:predictive}: $\boldtheta_{M+1} = \argmax_{\boldtheta} \mu_f(\boldtheta)$. However, with only a limited amount of observations, it is likely that the model does not capture the full range of variations in the figure of merit and this approach is likely to end up in a local minima. This can be seen for example in Fig.\ref{bofig}(a) where the model misses one of the peak.  Alternatively an explorative strategy would consist on taking measurements where the uncertainty in the model is maximal. This can be done by choosing $\boldtheta_{M+1} = \argmax_{\boldtheta} \sigma_f(\boldtheta)$, where the standard deviation $\sigma_f(\boldtheta)$ quantifies the uncertainty in the model. Balancing these two objectives, which are finding the location of the maximum and also exploring region of the parameter space where not enough observations have been made, is often referred as the \emph{exploration-exploitation} trade-off. In BO this trade-off is dealt with by introducing an \emph{acquisition function} $\alpha(\boldtheta)$ aiming at capturing these two aspects; and the choice of the next parameter is taken such that: 
\begin{equation}\label{eq:bo:acq:max}
	\boldtheta_{M+1} = \argmax_{\boldtheta} \alpha(\boldtheta).
\end{equation}
Among the most popular acquisition functions we recall the definition of the Upper Confidence Bound (UCB) function already provided in the main text:
\begin{equation}\label{eq:bo:acq:ucb}
	\alpha_{UCB}(\boldtheta) = \mu_f(\boldtheta) + k \sigma_f(\boldtheta),
\end{equation}
where the positive scalar $k$ balances the bias toward exploration (for high value of k) or exploitation (small values of k). This acquisition function was used to produce the graphs in the bottom panels of Fig.\ref{bofig}. In Fig.\ref{bofig}(a-b) a value of $k=4$ was used, while in \ref{bofig}(c) in the last iteration of the optimization it was taken to be $k=0$ in order to show the optimal parameter found according to the model.

Another acquisition function often used is the \emph{Expected Improvement} (EI) function, where the improvement, defined with regards to the best observation of the figure of merit obtained so far $y_{max}$, is averaged under the predictive distribution:
\begin{equation}\label{eq:bo:acq:ei}
	\alpha_{EI}(\boldtheta) = \int  max(0, f(\boldtheta) - y_{max}) p(f(\boldtheta)|\mathbf{y})df(\boldtheta),
\end{equation}
which has analytic solution \cite{snoek2012practical}. The choice of the acquisition function can have in some cases significant impact on the final results and both the UCB and EI functions were considered in this study. In practice finding the maximum of these acquisition functions is often performed by gradient ascent (possibly repeated over several random starting parameters). For both the UCB and EI, the gradients of the acquisition function with regards to the parameters have analytical expressions and can be efficiently evaluated. These two acquisition functions and the routines for finding their maxima are implemented in most of the BO libraries such as \cite{gpyopt2016}.

\section{Bayesian optimization: implementation details}\label{sec:appendix:impl}
The specific details of the BO runs used to obtain the results presented in Sec.\ref{orderorder} are given here.
\subsection{SF - MI transition}\label{sec:appendix:impl_sfmi}
For the SF-MI transition example, all the optimizations shown in Fig.~\ref{figdiffmethods} were performed with the same configuration. In each case, $N_{init}=100$ observations were initially taken for randomly chosen parameters followed by $M=1750$ iterations of BO. To decrease the computational burden associated to this large number of iterations we found it efficient not to update the hyper-parameters (noise and parameters of the kernel functions) at each iteration but rather every $10$ iterations. This resulted in a saving of almost $40\%$ in computational time without affecting the quality of the convergence results. For the choice of the next parameter to probe the UCB acquisition function in Eq.\ref{eq:bo:acq:ucb} was used with a value of $k$ linearly decreasing from $k(1)=5$ at the first iteration to $k(1750)=0$ at the end of the optimization. For this problem this acquisition function performed better than the EI one. Finally other types of kernels \cite{williams2006gaussian} were compared and we found that the \emph{Mat\'{e}rn 5/2} kernel performed almost always better than the \emph{Radial Basis Function (RBF)} kernel which is also a popular choice \cite{wigley2016fast}.

\subsection{Rydberg crystalline states}\label{sec:appendix:impl_rydberg}
For the creation of Rydberg crystalline states, the optimization problem was found to be simpler and the number of initial observations and iterations was significantly smaller than in the previous case. This can be attributed, in part, to the smaller dimensionality of the parameters vector. For Fig.~\ref{Rydprot} BO routines were initialized with $N_{init}=24$ observations followed by $M=10$ iterations. In the noisy case presented in Fig.~\ref{latticedefect} a larger number of iterations was needed: all the optimizations were run starting with $N_{init}=6$ initial observations and $M=50$ iterations. The EI acquisition function was used for all the optimizations included in this part.
\end{document}